\documentclass[superscriptaddress,
twocolumn,
showkeys, preprintnumbers,
floatfix,]{revtex4-2}

\usepackage{dynlearn}

\theoremstyle{definition}

\newcommand{\ourTitle}{Whales in Space:\\
Experiencing Aquatic Animals in Their Natural Place\\
with the Hydroambiphone
}

\begin{document}

\title{\ourTitle}

\author{\href{http://csc.ucdavis.edu/~chaos}{James P. Crutchfield}}
\email{chaos@ucdavis.edu}
\affiliation{\href{http://csc.ucdavis.edu}{Complexity Sciences Center},
Physics and Astronomy Department\\
University of California, Davis, California 95616}

\author{\href{http://csc.ucdavis.edu/~ajurgens/}{David D. Dunn}}
\email{artscilab@gmail.com}
\affiliation{\href{http://artscilab.com}{Art \& Science Laboratory}\\
613 C Street, Davis, California 95616}

\author{\href{http://csc.ucdavis.edu/~ajurgens/}{Alexandra M. Jurgens}}
\email{amjurgens@ucdavis.edu}
\affiliation{\href{https://www.inria.fr/en/inria-centre-university-bordeaux}{
Inria Centre}, University of Bordeaux, France}

\date{\today}
\bibliographystyle{unsrt}

\begin{abstract}
Recording the undersea three-dimensional bioacoustic sound field in real-time 
promises major benefits to marine behavior studies. We describe a novel
hydrophone array---the hydroambiphone (HAP)---that adapts ambisonic
spatial-audio theory to sound propagation in ocean waters to realize many of
these benefits through spatial localization and acoustic immersion. Deploying
it to monitor the humpback whales (\emph{Megaptera novaeangliae}) of southeast
Alaska demonstrates that HAP recording provides a qualitatively-improved
experience of their undersea behaviors; revealing, for example, new aspects of
social coordination during bubble-net feeding. On the practical side,
spatialized hydrophone recording greatly reduces post-field analytical and
computational challenges---such as the ``cocktail party problem'' of
distinguishing single sources in a complicated and crowded auditory
environment---that are common to field recordings.  On the scientific side,
comparing the HAP's capabilities to single-hydrophone and nonspatialized
recordings yields new insights into the spatial information that allows
animals to thrive in complex acoustic environments. Spatialized
bioacoustics markedly improves access to the humpbacks' undersea acoustic
environment and expands our appreciation of their rich vocal lives.
\end{abstract}


\preprint{bioRxiv 2023.12.24.XXXXX}
\preprint{arXiv 2312.XXXXX}

\maketitle

\section{Introduction}
\label{sec:introduction}

Marine mammals spend the bulk of their active lives submerged beneath the sea
surface. Given the relatively poor propagation of light compared to sound in
the ocean depths, the world of these animals is primarily acoustic.  These
factors greatly complicate relying solely on surface observations to address
the full diversity of their behaviors. Fortunately, in the last decade or so
scientists demonstrated the substantial benefit of undersea, comprehensive
tracking with, for example, skillful attachment of digital devices that monitor
animal behavior via sensors that record video, sound, location, depth,
pressure, temperature, and the like \cite{Gold17a}. The following describes
complementary benefits that come from recording the underwater
three-dimensional bioacoustic sound field in real-time.

\begin{table*}
\begin{tabular}{|c|l|c|c|c|c|}
\hline
\textbf{Medium} & \textbf{Density} & \textbf{Bulk Modulus} &
\textbf{Sound Velocity} & \textbf{Wavelength} &
\textbf{Wavelength} \\
                & \textbf{(kg/m$^3$)} & \textbf{(Pa)} &
\textbf{(m/s)} **** & \textbf{$100$ Hz (m)} **** &
\textbf{$1000$ Hz (m)} **** \\
\hline
Air & 1.225 *& \(1.42 \times 10^5\) & 343  & 3.43 & 0.343 \\
\hline
Fresh Water & 1000 ** & \(2.15 \times 10^9\) & 1482  & 14.82  & 1.482
 \\
\hline
Seawater & 1025 *** & \(2.29 \times 10^9\) & 1500 & 15 & 1.5 \\
\hline
\end{tabular}
\caption{Sound propagation differences in air and water:
	(i) Medium density, (ii) medium bulk modulus, (iii) sound velocity, and (iv)
	sound wavelengths at two different frequencies. Unlisted, but important is
	sound dispersion: the range of frequency-dependent velocities is markedly
	large in water.
*Standard atmospheric conditions (0 C or 32 F, at sea level).
	**Standard atmospheric pressure (1 atm.).
	*** At sea surface.
	**** Room temp (20 C, 68 F).
	}
\label{tab:Sound}
\end{table*}

\subsection{Whale Bioacoustics}

Sound propagation in water differs from that in air: sound travels five times
faster in water than in air, acoustic waves in water propagate with much less
dissipation, and different frequencies travel at different speeds. (See Table
\ref{tab:Sound}.) These phenomena make undersea sound markedly more complex to
analyze, understand, and harness. They complicate directly monitoring and
interpreting sound in the ocean. That said, these properties also mean there is
additional information available in ocean acoustic waves to be harnessed for
environmental sensing and for communication. (See App.
\ref{app:OceanBioacoustics}.)

To begin to address these challenges, we applied spatial bioacoustics to
monitor humpback whales (\emph{Megaptera novaeangliae}) of southeast Alaska,
demonstrating that it markedly improves understanding their undersea behaviors.
As one example, the following describes how acoustic spatialization revealed
previously unreported aspects of social coordination during bubble-net feeding
\cite{Crut23c}.

Cetaceans exhibit compelling evidence for advanced intentional behaviors and
conscious awareness through their raw intelligence, song generation
\cite{Payn83a,Payn00a} and sharing \cite{Garl13a,Garl17a}, communication and
interactions with their own and other species \cite{Smit08a,Chol18a}, and
empathy (concern for others' well-being) \cite{Pitm17a}. Humpback whales, in
addition, are known to be very vocal and social \cite{Payn95a}.

Evolving over a time span ten times that of humans, cetaceans developed tools
(socially-coordinated bubble-net feeding by humpbacks) and region- (and
possibly hemisphere-) spanning ocean-acoustic communication networks
\cite{Payn71a}. Over the last half century humpback whales, in particular,
became known for their active vocalizations. These fall into two categories:
One comprised of extended \emph{songs} (minutes to hours), emitted
predominantly by males; the other \emph{social calls}, short vocalizations
(lasting seconds) that occur in animal interactions and are produced by both
males and females \cite{Payn83a}.

Song function is still largely a mystery. Since songs are predominantly
produced by males, historically they have been interpreted as facilitating mate
choice and so playing a role in sexual selection
\cite{Wils00a,Gloc83a,Herm16a}. However, more recent results suggest that
humpback songs are not so much about reproductive fitness, rather they may
``reveal the precise locations and movements of singers from long distances and
may enhance the effectiveness of [acoustic] units as sonar signals''
\cite{Merc22a}. That is, vocalization is important to navigation which is
central to the very long seasonal migrations (1000s km) of humpback whales.

These debates highlight the need to discover patterns and the information
contained in animal vocalizations and to place these tasks at the center of
monitoring and interpreting cetacean behavior. This, in turn, calls for more
study and new acoustic instrumentation that reveals spatial aspects of whale
vocalizations. And, this suggests exploring new concepts of patterns and
structure in data as developed in the mathematical theory of causal statistical
inference \cite{Crut12a,Loom21b,Loom22a} and algorithmic approaches from modern
machine learning and AI (ML/AI) \cite{Berm19a}.

Both the foundational theory and ML/AI methods require substantial datasets and
in the latter case typically require labeling by human experts. We show
that new kinds of data from new kinds of instrumentation can be equally or more
important to interpreting acoustic data than exploiting massive data, algorithm
advances, and huge computational resources.

The principal reason to pursue innovations in instrumentation is that acoustic
spatialization disambiguates the locations of animal vocalizations, markedly
improving access to the humpbacks' undersea acoustic environment and expanding
our appreciation of their rich vocal lives. In addition, spatialized hydrophone
recording greatly reduces the post-field analytical and computational
challenges when confronted by many vocalizers and multiple additional sound
sources, as commonly occurs in field recordings. In contrast to these
well-known difficulties, it is important to keep in mind that human observers
and apparently marine mammals are not prohibitively challenged by
multiple-source ambiguity---the so-called ``cocktail party problem''.

The following introduces a novel hydrophone array---the
\emph{hydroambiphone}---that adapts ambisonic spatial-audio theory to sound
propagation in ocean waters to largely alleviate such problems, while providing
(i) clues to the kinds of information that allow animals to thrive in complex
acoustic environments and (ii) markedly more representative experiences of
their acoustic world.

\subsection{Undersea Ambisonics}

To explore the complexities of the undersea acoustic environment and circumvent
these problems, we adapted the ambisonic theory of spatial acoustics
\cite{Gerz75a,Zott19a}. The theory formalizes the representation of an ambient
sound field systematically in terms of three-dimensional spherical
harmonics---the solutions of the wave equation that describes sound propagation
in a medium.  The theory gives an exact representation of a sound field
centered at a given point in space in terms of a systematic approximation at an
infinite number of ``orders''. In practice one can only go to finite-order
approximation. The higher the order, though, the better the spatial resolution
of the approximation, but the number of required transducers grows rapidly with
order. For this reason, real applications generally use low-order
approximations. Ambisonic theory applies to both recording and playback of
three-dimensional spatial sound fields.

In-air ambisonic recording arrays use directional microphones---cardioid or
super-cardioid, for example. They are placed and oriented to completely cover a
sound field with non-overlapping regions. This is not possible in water, since
the available transducers---hydrophones---are omnidirectional. To address this
our implementation uses a first-order ambisonic approximation consisting of
four omnidirectional hydrophones mounted on the surface of a 12-inch diameter
hollow sphere. Given the corrosive nature of seawater the sphere was stainless
steel. It also provided a high specific acoustic impedance to isolate
hydrophones from acoustic waves propagating from directions opposite each
hydrophone. Thus, the spherical shape provides a kind of ``acoustic shadowing''
that improved the directivity of each hydrophone, whose individual sensitivity
is otherwise omnidirectional, as noted.

The following demonstrates how to adapt ambisonics to the ocean acoustic
environment and proves out the HAP as an effective marine bioacoustic
instrument. The purpose in this is two-fold: (i) introduce the HAP
and describe its deployment and (ii) recount several novel scientific results
from our August 2023 voyage that established its use as a viable marine science
instrument.

To these ends, we briefly layout HAP design (Sec.
\ref{sec:SpatialBioacoustics}),
calibration and performance (Sec.
\ref{sec:HAPPerformance}), spatial audio signal processing (App.
\ref{app:ProcessChain}), and listening (App. \ref{app:Listening}). We do cite,
however, companion technical descriptions that go into considerably more
detail. The main focus here are the scientific results that came from using the
HAP's spatial audio: (i) access to a new immersive experience of the humpback
acoustic Umwelt, (ii) novel acoustic coordination during bubble-net feeding,
(iii) dynamics of group breaching, (iv) HAP high acoustic sensitivity, (v)
undersea infrasonics, and (vi) undersea noise in the Inside Passage.

\subsection{Overview}

The following recounts the results of deploying the HAP during a 300 mile
voyage along the Inside Passage of southeast Alaska 18 August to 2 September
2023, transiting from Juneau to Ketchikan aboard the M/Y Blue Pearl
(Don and Denise Bermant, owner/operators).

First, we review ambisonic spatial audio theory and then move on to describe
how this is adapted to the design of a hydrophone array that records the
undersea three-dimensional sound field. We comment on several design aspects,
its testing and calibration, and required digital audio recording and playback
for listening. We summarize the HAP's performance in Sec.
\ref{sec:PerfConfounds}.

We then review a selection of new undersea bioacoustic phenomena, most related
to humpback whale behaviors and vocalizations, but also recount several notable
purely ocean-acoustic observations. This includes a brief discussion of the
present challenges.
We end
by drawing out a number of conclusions that range from emphasizing the novelty
of the whale bioacoustics identified to suggesting future prospects for greatly
improved HAP systems and the possible innovations in marine bioacoustics.

\section{Capturing Spatial Bioacoustics}
\label{sec:SpatialBioacoustics}

A sound field is a three-dimensional organization of acoustic energy sustained
by oscillatory motions of a medium, such as air or water. When a sound source
activates, the field agitation propagates in all directions away from the
source in spatiotemporal patterns governed by the wave equation:
$\ddot{\vec{u}} = c^2 \nabla^2 \vec{u}$, where $c$ is the speed of sound and
$\vec{u} (x,y,z)$ is local state of the medium. The result is that monitoring
the instantaneous state of a time-dependent sound field requires measuring and
then recording a (vector valued) function $\vec{u}$ throughout a
three-dimensional volume.

When we listen to an audio recording of a musical performance, our ears are
presented with a sound field that is generated by a particular spatial
configuration of voltage-to-pressure transducers (loudspeakers) that are driven
by signals picked up by artfully-placed individual sound-to-voltage transducers
(microphones). It is one of the sound recording engineer's primary
responsibilities to determine the types of microphone and their placement---a
contact small-diaphragm condenser mic near sound hole of the acoustic guitar, a
dynamic mic near the singer's mouth, a unidirectional cardioid close to and
facing the drums---to best capture and then reproduce the musical experience.

Unfortunately, the selected mic placements anchor the recorded sound signals to those specific locations. Once recorded, those locations cannot be changed.
Post-recording, there is little variation available to acoustically explore,
since so much of the original ambient 3D sound field $\vec{u}$ information is
lost in a collection of spatially-local recordings. Moreover, careful and
extensive mic placements are typically not feasible nor is the resulting sound
reproduction adequate for natural sound fields.

\subsection{Ambisonics}

An alternative way to capture a 3D sound field is afforded by adapting
ambisonic (spatial audio) theory \cite{Gerz75a,Zott19a}.
Its benefits include (i) listener-centric representation of the
sound field (rather than a set of mic-centered signals), (ii) flexible
post-recording signal processing, such as panning, zoom, rotation, and beam
forming, (iii) flexible post-recording encodings to an unrestricted array of
playback systems---from monophonic and stereo to modern full immersion systems
with dozens or even hundreds of loudspeakers. As we will show, these benefits
are key to processing and interpreting vocalizations of animals, especially
those that move in three dimensions.

Ambisonics does this by expanding the sound field in spherical harmonics
centered around the listening point (or \emph{sweet spot}). These spherical
harmonics are the solutions to the 3D wave equation above and give a
mathematically-consistent and systematic approximation basis for representing
the ambient sound field up to a given \emph{ambisonic order}.

\subsection{Hydroambiphone}
\label{sec:HAP}

\begin{figure}[t]
\centering
\includegraphics[width=.8\columnwidth]{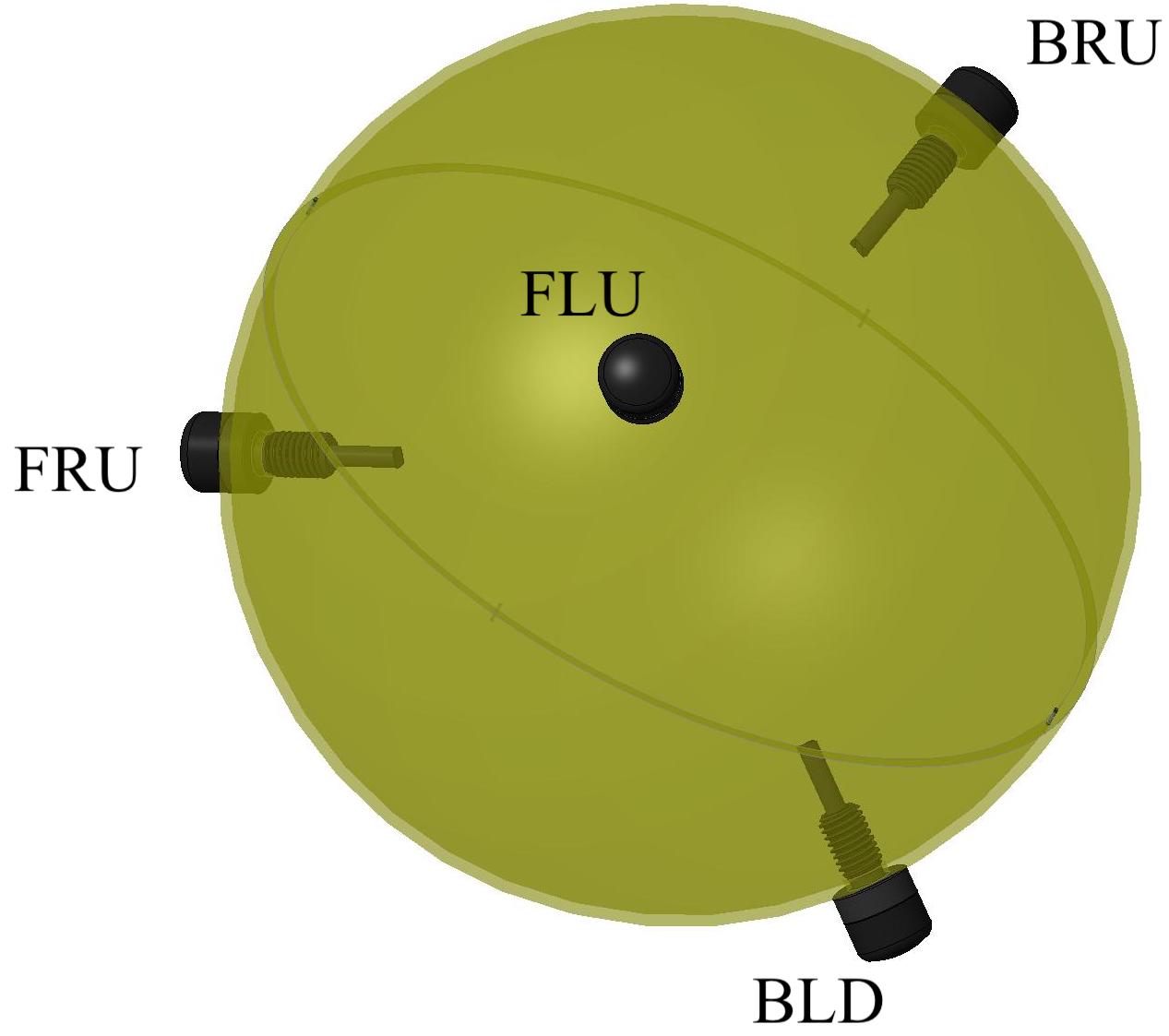}
\caption{Hydroambiphone with hydrophone placements on the surface of a hollow
	sphere: (i) Front-Left-Up (FLU) (ii) Front-Right-Down (FRD), (iii)
	Back-Left-Down (BLD), and (iv) Back-Right-Up (BLU).
    }
\label{fig:HAP}
\end{figure}

Ambisonic recording produces a time series of measurements each of which is
vector of acoustic signals. The dimension of the vector grows quadratically
with the approximation order. Thus, implementing an ambisonic transducer
entails practical trade-offs between, for example, the expense and cabling of
multiple transducers, large data storage requirements, and computational
complexity and computing resources required to process long-duration vector
signals---in real-time and off-line. Balancing these, we selected a first-order
ambisonic (FOA) array consisting of four hydrophones mounted on the surface of
12 inch-diameter hollow sphere. To properly ``shadow'' each hydrophone from
sounds arriving from opposite directions (and so improve directionality) the
sphere was 2mm thick stainless steel, which has a high specific acoustic
impedance. (See Apps.
\ref{app:F8nPro} and \ref{app:RecordingSystem} and Fig. \ref{fig:HAP}.)

\subsection{Ambisonic Recording and Signal Processing}

The HAP acoustic signals were captured by a multi-channel digital audio
recorder (Zoom F8n Pro) to high-capacity, high-speed SD cards.  Each of the
HAP's four channels was sampled at 24 bit resolution at a rate of 44.1 kHz in
the recorder's Ambisonic mode. After a day's collection of recording sessions,
the audio data was transferred to a laptop and also to an additional backup
disk. Figure \ref{fig:RecordingSystem} shows the complete first-order Ambisonic
recording and observation system.

The spatial audio processing chain consisted of a specific series of stages:
\begin{enumerate}
      \setlength{\topsep}{-3pt}
      \setlength{\itemsep}{-3pt}
      \setlength{\parsep}{-3pt}
\item Cleaning each HAP Ambisonic A mono channel to attenuate interference
	from vessel operating subsystems, such as sonar, and also from surface
	wave noise.
\item Converting the Ambisonic A Format vector time series to Ambisonic B
	(Fuma) format, accounting for the physical properties of sound in water and
	the HAP's design.
\item Beamforming the B format vector signal to increase the directionality of
	the encoded ambisonic signals.
\item Encoding the beamformed Fuma signal to one or another playback
	configuration. These included stereo, binaural, a custom 6.1 Hexagonal F
	encoding, and several Dolby Atmos configurations ranging from a half-dozen
	to two dozen speakers.
\end{enumerate}
For details see App. \ref{app:ProcessChain}.

The listening experience of spatialized audio depends strongly on the playback
system. Available audio formats and loudspeaker configurations offer very
different degrees of spatialization and acoustic immersion. Headphone listening
via the binaural encodings is convenient and captures a notable amount of the
spatial audio, given that it is being heard over two-channel stereo headphones.
In contrast, Hexagonal F surround presents a richer sound environment with much
improved source localization. Our custom-built Hexagonal F 6.1 surround
playback systems consist of a subwoofer and six co-planar satellite speakers
(located at head level and centered around the sweet spot where the listener
sits). Listening via the Hexagonal F system is necessary to discriminate
sources in complex acoustic scenes and/or with low-level or distant acoustic
sources.

\section{HAP Performance}
\label{sec:HAPPerformance}

Our main message here is that the HAP worked to capture the undersea 3D
acoustic sound field. In point of fact, its performance greatly exceeded our
original predictions. That said, the field deployments also revealed a number of
challenges and so opportunities for improvement.

Before recounting the scientific results, we will describe the HAP's initial
field deployment on our August 2023 voyage and mention several practical issues
for effective use. Second, we will then describe the HAP's basic and spatial
acoustic performance, including calibration test and a number of experimental
confounds.

\subsection{Field Deployment}

Our field deployments of the HAP occurred over two weeks (Summer-Fall 2023) on
a 300 mile voyage through the Inside Passage of southeast Alaska, transiting
from Juneau to Ketchikan aboard the M/Y Blue Pearl, a 65' Fleming motor yacht.
Of note for supporting our field tests the vessel provided 115 V AC power for
the recording and playback equipment and device battery charging and Starlink
marine uplink to the Internet with 10 Mbps upload, 20-50 Mbps download, and
very low latency.

The HAP was supported by a stainless steel cable that relieved stress from the
four hydrophone signal cables. Given the 60' long cables the HAP was deployed
from 15' - 50' depths. It was particularly important to carefully track the
vessel's position and orientation with respect to current, wind, drift, and
surface waves. Without monitoring these, the strong Alaskan currents would
sweep the HAP under the vessel and wave slap against the vessel hull could
become a distracting source of acoustic noise.

\subsection{Performance and Acoustic Confounds}
\label{sec:PerfConfounds}

Due to the focus here on the HAP's practical use as an instrument for marine
animal behavior, we briefly outline its performance. Detailed quantitative
performance measurements will be reported elsewhere.

As far as detecting vocalizations from aquatic animals, the relevant aspects of
the HAP's performance include: (i) high sensitivity, (ii) long distance
detection of sound sources (including marine mammals and human-generated
sounds), and (iii) good directivity resolution.

The high sensitivity manifested itself by the HAP's ability to detect extremely
low-level sound sources---including both sources that originated close to the
vessel, such as wave noise from the vessel's hull, and sources, such as engine
noise, from other vessels over a dozen miles away (well beyond the horizon and
simply not visible).

Initially, the HAP's high sensitivity seemed detrimental to the recordings as
it added many nonbiological sources not of direct interest. For example, one
persistent problem in the Inside Passage was the (even very distant) transit of
cruise liners. Their noise amplitudes at the HAP were so large that we typically
aborted recording sessions as the animal vocalizations could be entirely masked
and the incoming signals were painful for human listeners. Subsequently,
however, we were able to remove and attenuate these noise sources during
post-field processing largely due to the Ambisonic benefits noted above.

An important parameter of the HAP was its directional sensitivity: how close
(in solid angle) can independent sources be and still be distinguished.
Pre-field tests in our lab tanks indicated being able to distinguish sources at
90 or more degrees of solid angle---octants on a sphere centered at the HAP.
However, field tests with the vessel's tender circulating at various radii
indicated 45 degrees of solid angle distinguishability.

\section{Undersea Acoustics Revealed by the HAP}

On our two week voyage and over dozens of recording sessions, the HAP proved
itself to be a remarkably sensitive, flexible, and easy-to use instrument for
capturing the undersea three-dimensional acoustic world. This led to a number
of insights and discoveries, ranging from diverse animal behaviors and
vocalizations to providing spectacular undersea immersive audio listening.
These included: a wholly new appreciation of the whales'
acoustic \emph{Umwelt}; humpback pairs harmonizing their social calls during
collective bubble-net feeding; the HAP's high acoustic sensitivity over long
baselines; the dynamic interplay of humpback group breaching; extensive
recordings of undersea noise sources; and the detection of undersea infrasonics
(sound frequencies under 20-30 Hz). We briefly outline each, leaving fuller
accounts to follow-up companion articles.

\subsection{Intangibles made Present}

As observers listening to the HAP stereo signal in real-time, were daily
surprised by the high degree of acoustic activity. Due to this, the voyage
resulted in over 70 HAP recordings (from 10s of minutes to an hour in length
each), in toto representing many dozens of hours of observation and recording.
By removing silent sections in post-field processing, we compiled the
collection of recordings into a single immersive audio file two hours and
forty-seven minutes long that highlights the huge diversity of humpback
vocalizations.

Overall, the compilation provides a thorough-going immersion into the acoustic
environment experienced by southeast Alaskan humpbacks. While listening
on-board in real-time was surprising in a number of ways, the compilation
recording, being a concentrated presentation of selected examples, led to even
more insights. We were regularly surprised at the number of animals around us.
This activity was far in excess of the numbers we expected from our visual
surface observations. Another general observation was the shear diversity of
humpback social calls and the regularity of apparent acoustic communication. It
became clear that their vocal activity was quite high. This stands in contrast
to the oft-expressed belief that, due to their preoccupation with feeding,
during the summer months in southeast Alaska humpbacks are markedly less vocal
than during the winter mating season at low latitudes in Hawaii.

The overall take-away message is a holistically and deeply enriched experience
of the humpbacks' spatial acoustic world--their \emph{Umwelt}
\cite{Uexk92a}---that is difficult to express in the written words.  To
demonstrate, we are making public a short 20 minute excerpt from the
compilation that presents several highlights, available
\href{http://demon.csc.ucdavis.edu/~chaos/share/www/WorldWideWhale/World_Wide_Whale.html}{World
Wide Whale}.

\subsection{Humpback Harmonies}

Spatializing the social calls used by humpbacks during bubble-net feeding
revealed previously-unreported vocal coordination of different individuals.
Specifically, by listening to the immersive audio encoding (binaural but
especially the Hexagonal F surround system) and closely inspecting spectrograms
of the section of bubble-net feeding, revealed that a second humpback joins the
main feeding call just before the end of the coordinated feeding activity.

\begin{figure}[t]
\centering
\includegraphics[width=.48\textwidth]{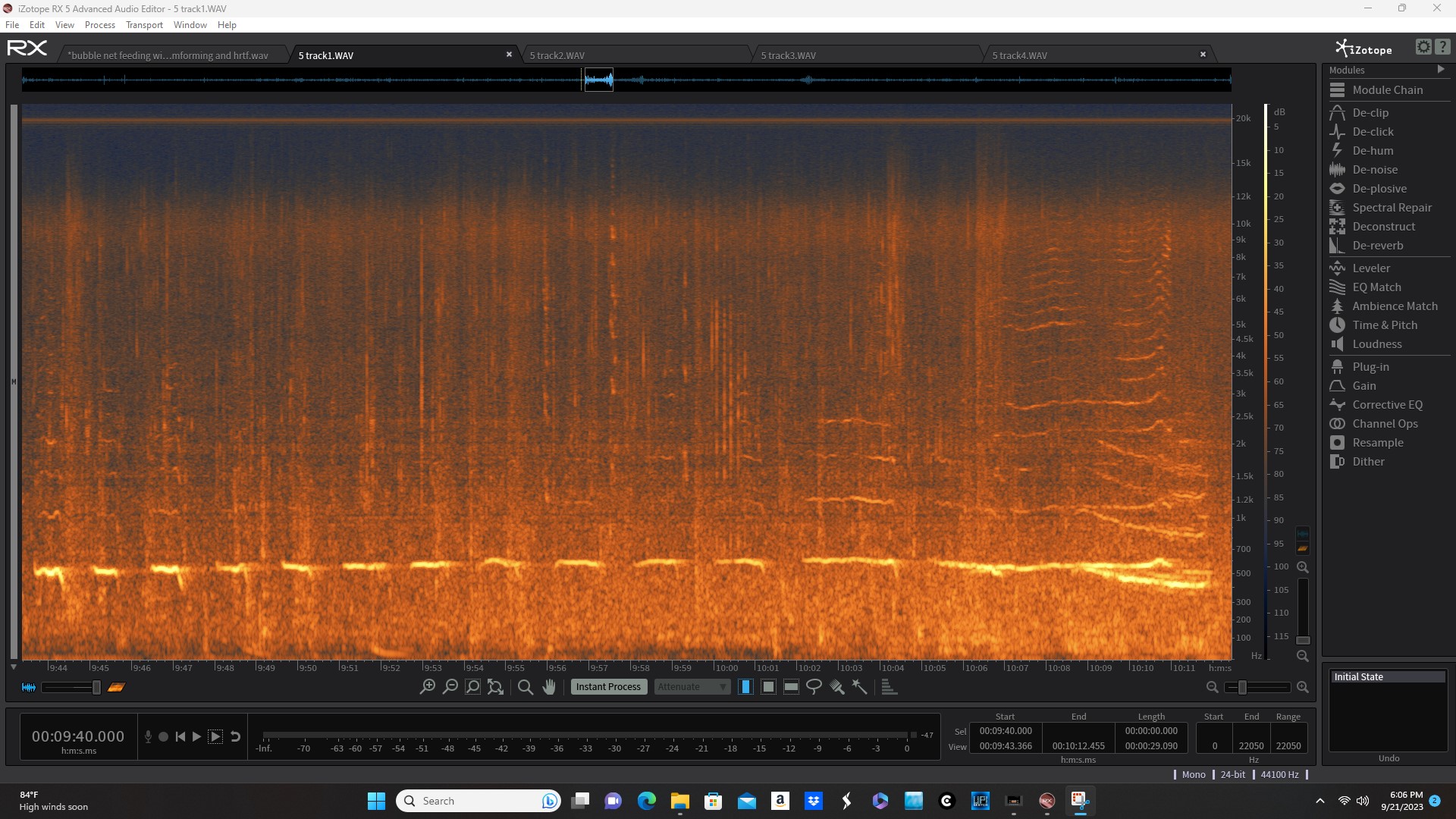}
\caption{Spectrogram of a group bubble-net feeding session in which
	a second whale joins the existing main coordination feeding call vocalized
	by a different animal. Spectrogram calculated from the unaltered HAP
	Ambisonic A recording on HAP channel 1 only. Note where the second whale
	joins and how the harmonics separate and go in two directions at the very
	end: one up in frequency and the other down.
	(Reproduced here from Ref. \cite{Crut23c} with permission.)
}
\label{fig:Harmony}
\end{figure}

The spectrogram analysis used the unaltered HAP Ambisonic A recordings
corresponding to original channel numbers. See Fig. \ref{fig:Harmony} for the
spectrogram of HAP channel 1. The spectrogram makes it clear when the second
whale vocally differentiates: the two animals' harmonics separate and go in two
opposite directions at the very end. Moreover, listening to the spatialized
recording makes it clear that the source of the vocalizations consists of two
distinct animals at different locations.

In addition, the recording reveals that initially, before the frequency
separation, the two animal's vocalizations start off synchronized at the same
frequency. Only then does one animal shift up in frequency (as is typical in
bubble-net coordination calls from lone individuals) while the other shifts
down. It also clear that the phase and amplitude differences between the
animals' calls lie mostly in the mid- and high-frequency ranges, as expected.
One concludes that as a vocal phenomenon these vocal coordinations are the
functional equivalent of human singers harmonizing.

We recorded four such harmonizing events. Fuller description, recording data,
and analysis are presented in Ref. \cite{Crut23c}. The latter also further
explores the benefit of the HAP's spatialization that reveals the two animals
are clearly vocalizing from two different locations.

\subsection{High Acoustic Sensitivity}

The HAP revealed itself to be surprisingly sensitive, for close and very
distance sound sources. For example, on board listeners often heard cruise
liners and small fishing boats (their outboard motors in particular), long
before they could be seen above the ocean horizon. This places these sound
sources more than about $8$ miles away from the vessel.

The high sensitivity at first seemed a burden, obscuring and even totally
masking sounds of interest. It certainly was while listening on board. While in
those situations the HAP's high sensitivity seemed a detriment, our ultimate
view is that the high sensitivity facilitates the HAP's (and our) probing
complex aspects of the 3D acoustic ocean environment in constructive
ways---ways that reveal new phenomena. Specifically, when we returned
post-field to the lab we developed the spatial audio processing chain described
in App. \ref{app:ProcessChain} to largely remove or attenuate many of the
``extraneous'' sound elements that the HAP's sensitivity introduced.

\subsection{Discussion}

Other follow-on companion works will report on additional notable results from
the HAP. These will include a spatial analysis of long periods of group
breaching. From these one can estimate aspects of their configuration and even
the local bathemetry. This then raises the question in a newly quantitative way
of the function of breaching. Extracting spatiotemporal information from the HAP
recordings in terms of contingency between breaches will allow one to probe if
breaching supports communication. Another notable observation, one that at
present is not explained and is counterintuitive, is repeated recording of
undersea infrasonics that were highly directional. The directionality goes
against conventional understanding of sound propagation since low frequencies
are not associated with having a direction. Finally, over the voyage we came
into contact with many natural and anthropogenic sound sources. In this way,
the collection of HAP recordings can give insight into a number of undersea
noise sources. We believe that the spatial aspects of these noise sources
have much to contribute to our understanding of noise in the undersea
environment and to monitoring how they affect the animals there.

\subsection{Related Work}

Several independent efforts have attempted to develop devices for undersea
spatial audio recording over the last decades. These range from developing
directional hydrophones to hydrophone arrays. For example, Sonotronics offers
their
\href{https://www.sonotronics.com/model-dh-5-directional-hydrophone/}{Model
DH-5 Directional Hydrophone}. More relevant is an implementation for
stereophonic underwater sound recordings disclosed in
\href{https://patents.justia.com/patent/8509034}{US Patent \#8,509,034}.
Finally, open source software has been developed for real-time acoustic
detection and localization of cetaceans \cite{Gill09a}.

The most similar prior effort, though, is found in Refs.
\cite{Fari11a,Fari12a,Fari18a,Fari19a}. These implement an open-frame
four-hydrophone array in a tetrahedral configuration---that is, a first-order
ambisonic approximation.  Unfortunately, the spatial localization was not
strong, which appears largely due to the use of an open frame mounting for the
hydrophones.

To the best of our knowledge our reports here are the first successful
undersea spatial audio recording and the first used to successfully
study cetacean behavior.

\section{Conclusion}

Environmental and behavioral marine science and technology have changed
immeasurably since the early days of the first appreciation of how sound
propagates in the ocean and of who and what are producing those sounds.
Certainly, hydrophones and ancillary signal processing have advanced
substantially. These improvements promise to greatly enrich our understanding
of the undersea world and its inhabitants.

We introduced a modest but accessible implementation of undersea spatial
acoustics that uses inexpensive commercial-off-the-shelf (COTS) hardware and
open-source software. The net functionality of the recording system allowed for
real-time spatialization of aquatic animal vocalizations. We believe the HAP is
a new tool for marine biology that promises to greatly expand the human
appreciation of the three-dimensional acoustic world of marine animals.

Our reports of the successful proof-of-concept deployment also suggests
substantial improvements. So, one can look forward to future implementations
that provide increasingly higher quality and improved sensitivity and higher
resolution localized spatialized sound, all available in real-time.

The operant question now is how to actively shape our future understanding of
whale communication in the wild. The results suggest a coming era of citizen
marine social science. The hydroambiphone is straightforward to construct at
moderate cost and so gives a practical path to a new era of listening to the
voices of the deep.

\section*{Acknowledgments}
\label{sec:acknowledgments}
\vspace{-0.2in}

The authors are deeply indebted to Don and Denise Bermant for their generous
tour of the Inside Passage, southeast Alaska, on the M/Y Blue Pearl (Vancouver,
British Columbia) and their skillful navigation, marine insights, and patience,
as well as camaraderie. All were key to the voyage's success. They also thank
Anwyl MacDonald for 3D modeling, Dave Hemer (UC Davis) for skillful machining,
Robb Nichols (Aquarian Audio \& Scientific) for advice and early hydrophone
access, and Gary Elko and Jens Meyer (MH Acoustics) for advice on in-air
ambisonics. They also thank Kelly Finn for surface observations and videography
and insights on the appropriateness of contemporary animal-behavior observation
protocols for revealing marine mammal behavior.

{\bf Author contributions}:
J.P.C. designed and built the HAP underwater hydrophone recording array and
underwater acoustic broadcast system with the expert advice of D.D.D. J.P.C.
and D.D.D. conducted the field recording experiments and calibration of the
recording and playback subsystems. A.M.J. was responsible for field photography
and observations of surface behavior. These efforts were conducted using
personal and University of California equipment. J.P.C., D.D.D., and A.M.J
contributed to data acquisition, analysis, and interpretation. They also
contributed to all aspects of manuscript writing and production.

{\bf Funding}: The authors' efforts were supported by, or in part by, Templeton
World Charity Foundation (TWCF) grant TWCF0570 and Foundational Questions
Institute and Fetzer Franklin Fund grant FQXI-RFP-CPW-2007 both to the
University of California, Davis (Lead P.I. J. P. Crutchfield), and the Art and
Science Laboratory. The opinions expressed in this publication are those of the
authors and do not necessarily reflect the views of the Templeton World Charity
Foundation, Inc.

{\bf Competing Interests}: None declared.

{\bf Data and materials availability}: Data provided by the first author upon
reasonable request.

Harpex is a trademark of Harpex Audio GMBH.

\appendix

\section{Ocean Bioacoustics}
\label{app:OceanBioacoustics}

Sound propagation in water differs markedly and in key ways from propagation in air. Given human's innate sense and experience of sound in air, the differences need to be taken into account when interpreting the signals that hydrophones pick up.

First, the speed of sound in water is five times that in air: $1,500$ meters
per second compared to $340$ meters per second, respectively, owing to the
water medium being markedly denser than air. Practically, this leads to, for
example, echos as sounds bounce off the seabed. Since density increases with
depth, water depth is important and, of course, changes when changing
anchorages. This also means that sounds from distant sources can be detected.
For example, one is often surprised by the degree to which vessel noise is
heard and in some cases dominates the undersea soundscape, even if vessels are
not in sight. Commercial cruise liners are notable contributors to ocean noise
given their immense displacement (key to waves generated by their passing) and
massive engines.

Second, the precise nature of propagation in water is complicated by the fact
that sound velocity increases with water pressure (and so depth) and decreases
with water temperature and salinity.

Third, unlike sound in air, underwater sound at different frequencies
propagates at different speeds---this is referred to as frequency dispersion.
Thus, a distinct sound pulse detected at some distance loses its sharpness and
blurs out over a time period much longer than the original pulse.

Taken altogether, the effects of these dependencies have on propagation are
unlike those of our experience of sound in air. They often result in unusual
and counterintuitive sound phenomena. The physics underlying these effects are
nicely recounted in Ref. \cite{Payn95a}.

For example, the dependencies lead to a fascinating phenomenon of extremely
long-ranged detection of sound signals in the ocean. This is the \emph{Sofar
channel}. Due to the competing effects of pressure and temperature on sound
speed, there is a horizontal ``channel'' that conducts sounds like a waveguide:
signals within a certain frequency band bounce between a shallow ``ceiling''
(perhaps $10$s of meters in depth) and a ``floor'' ($100$s meters or more in
depth). The net result is that sound signals in the Sofar channel can propagate
very long distances---easily tens of kilometers or, depending on conditions, to
hundreds or thousands of kilometers.

One the one hand, these properties mean that undersea sound is markedly more
complex to analyze, understand, and harness. These complications add challenges
both to successful life undersea and to directly monitoring and interpreting
sound in the ocean. On the other hand, these properties also mean there is
additional information available in ocean acoustic waves---information that can be harnessed for environmental sensing and for communication.

Given the undersea is their environment and given their evolution over millions
of years, marine animals, such as whales, have accounted for and take advantage
of these ocean-acoustic properties. These features affect what they can
perceive, how they generate sound underwater, and how they communicate and
socialize. Undoubtedly, many aspects of their vocalizations are naturally
adapted.

Finally, these properties affect the acoustic signals one records via
hydrophones and so, too, how one interprets what one is hearing.

Properties that largely determine sound propagation speeds and wavelengths in
air and water are given in Table \ref{tab:Sound}.

\section{Digital Audio Recording}
\label{app:F8nPro}

The HAP acoustic signals were captured by a multi-channel digital audio
recorder (Zoom F8n Pro) to high capacity SD cards.

Each of the HAP's four channels were sampled at $24$-bit resolution at a rate
of $44.1$ kHz. After a day's collection of recording sessions, the audio data
was transferred to a laptop and also to an additional backup disk.

\begin{figure*}[t]
\centering
\includegraphics[width=.8\textwidth]{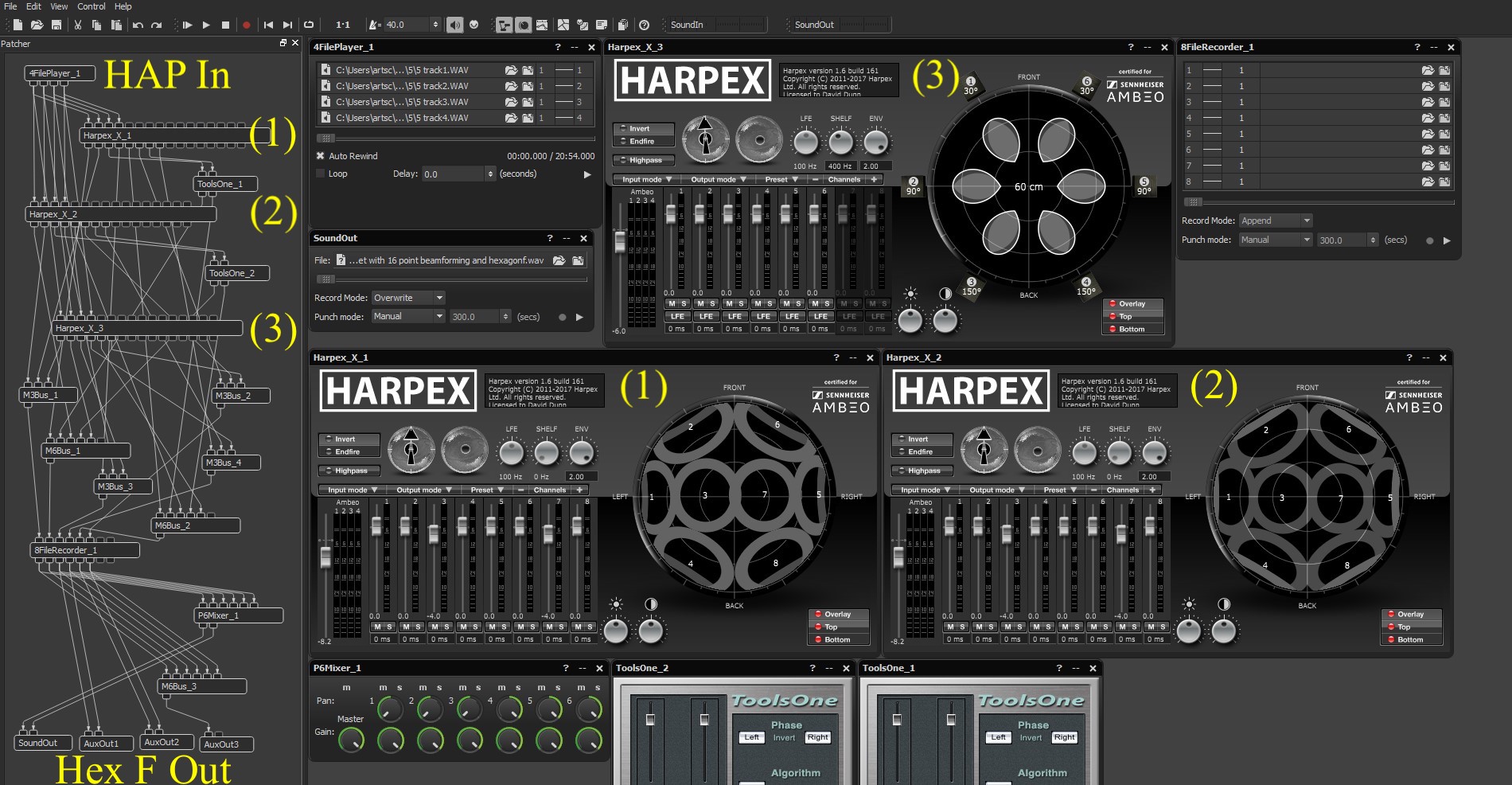}
\caption{Spatial Audio Processing Chain: Multi-Harpex configuration within
	AudioMulch VST host for processing HAP first-order (4 channel) ambisonic
	A-Format hydrophone signals (HAP In) to Hexagonal F 6.1 surround format (3),
	using beamforming (1) and (2) to increase the directionality of the decoded
	Ambisonic signals and to steer away extraneous sound sources.
    }
\label{fig:SpatialAudioProc}
\end{figure*}

\section{Data Acquisition and Recording System}
\label{app:RecordingSystem}

\begin{figure*}[t]
\centering
\includegraphics[width=.7\textwidth]{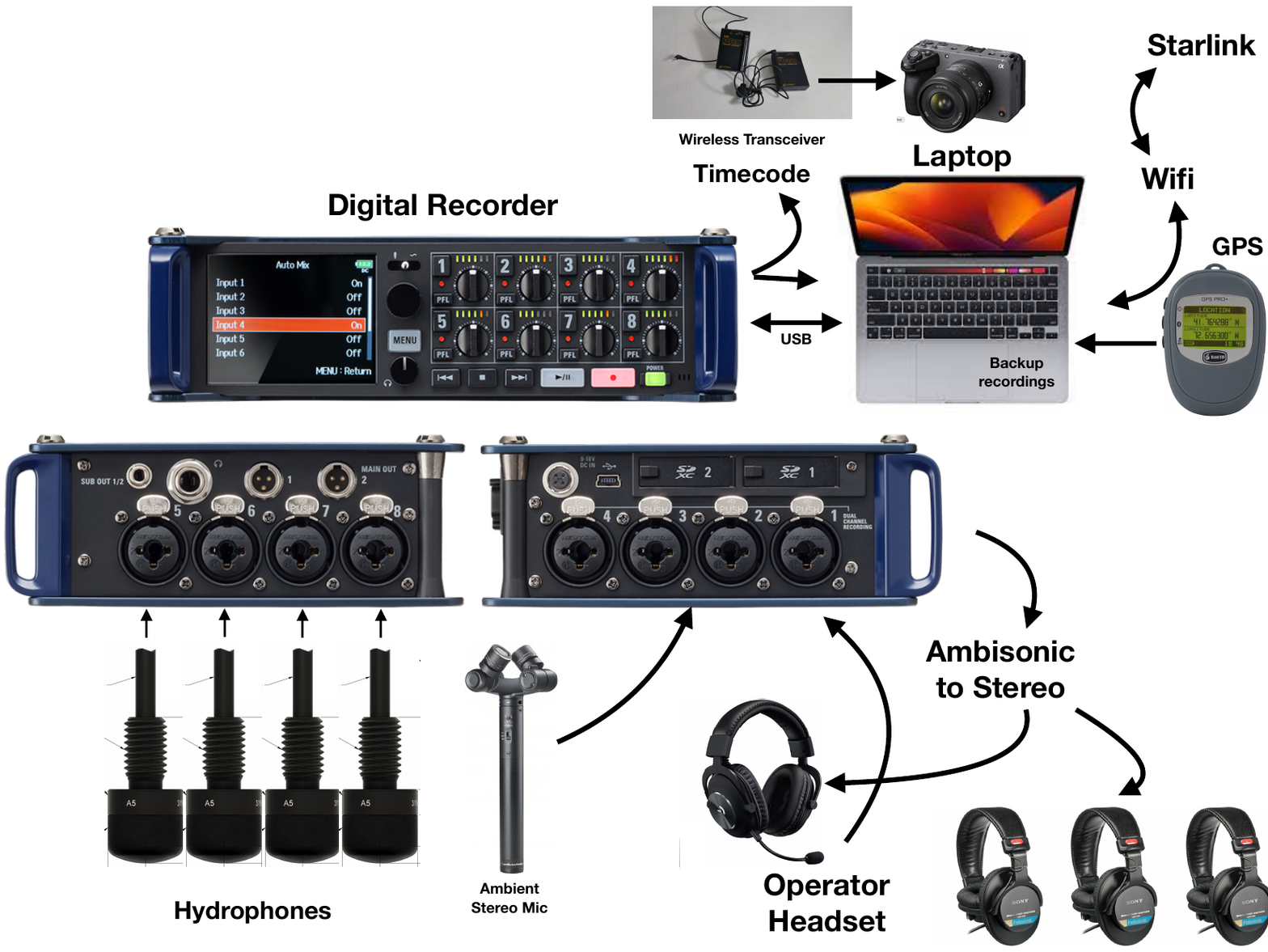}
\caption{First-order Ambisonic Recording and Observation System.
    }
\label{fig:RecordingSystem}
\end{figure*}

The complete data acquisition and recording system is shown in Fig.
\ref{fig:RecordingSystem}.

\section{Spatial Audio Signal Processing Chain}
\label{app:ProcessChain}

The following outlines, in the sequence used, the signal processing steps
developed to spatialize multichannel hydrophone recordings of underwater sound
sources, such as the humpback whales who are the focus here.
Section \ref{sec:HAP}
described the multichannel ambisonic hydrophone (HAP) recording
device that produced the 4-channel spatial audio signals---so-called Ambisonic
A Format. Familiarity with the HAP device and general ambisonic processing
theory \cite{Zott19a} is assumed.

\begin{enumerate}
      \setlength{\topsep}{-3pt}
      \setlength{\itemsep}{-3pt}
      \setlength{\parsep}{-3pt}
\item The raw Ambisonic A audio files were recorded as WAV Poly format with a
	\href{https://zoomcorp.com/en/us/field-recorders/field-recorders/f8n-pro/}{Zoom F8n Pro digital audio recorder}.
	These were then separated into four mono WAV files using the
	\href{https://www.sounddevices.com/product/wave-agent-software/}{Wave Agent
	(Sound Devices)} software application.
\item Each mono channel was separately ``cleaned'' in
\href{https://www.izotope.com/en/products/rx.html}{iZotope RX10} software to
	remove intermittent clicks and interference noise generated by the operating
	systems of the host sea vessel---i.e., sonar pinging, water filtration pump,
	navigation telemetry, and the like) and amplitude adjusted with identical
	settings applied to each Ambisonic A channel file.
\item The cleaned mono files were subsequently recombined into WAV Poly format
	and loaded into a channel of the digital audio workstation (DAW)
	\href{https://www.reaper.fm}{Reaper}.
	Within Reaper, the Sparta Array2sh VST plugin was used to convert the
	Ambisonic A files to Ambisonic B (Fuma) format while adjusting for the speed
	of sound in water (approximately five times faster than in air) and
	correcting for the specific transducer positions on the 12 inch diameter
	spherical housing of the HAP array. These adjustments were essential
	to account for the propagation of sound in the ocean environment and the
	relative arrival times of sound waves to the individual hydrophone positions
	as necessary to maintain the precise phase accuracy of the ambisonic
	encoding.
\item Once the audio data files were converted into Ambisonic B (Fuma) format,
	they could be encoded into the variety of binaural, surround, or speaker
	dome configurations that the ambisonic format facilitates. The
	\href{https://harpex.net}{Harpex X} VST plugin was used to perform this
	encoding function due to its flexible interface and direct support for a
	variety of binaural and surround configurations that can also be visually
	displayed in real-time as a useful two-dimensional mapping of
	three-dimensional space.
\item An additional useful feature of the Harpex X plugin is its support for
	creating ``beamforming'' adjustments to increase the directionality of the
	encoded ambisonic signals. Multiple instances of the Harpex X plugin can be
	activated simultaneously to create multiple combinations of beamformed
	signals. This can be used to increase the amplitude sensitivity of the
	overall system with regard to sounds occurring at extreme distances.
\item The VST plugin host \href{http://www.audiomulch.com}{Audiomulch} was used to simultaneously run multiple
	instances of Harpex X and to transparently mix the correlated output
	signals for assignment to various surround configurations or binaural
	output. A standard Head Related Transfer Function (HRTF) was used for
	conversion to binaural audio. In this case it was the widely used HRTF
	derived from the
	\href{https://www.neumann.com/en-en/products/microphones/ku-100/}{Neumann
	KU 100 Binaural Dummy Head microphone system}.
	See Fig. \ref{fig:SpatialAudioProc}.
\item All subsequent editing of the Ambisonic B or binaural audio
	files---prepared for public dissemination---was performed in either Reaper
	or \href{https://www.adobe.com/products/audition.html}{Adobe Audition}. No
	other equalization, filtering, signal processing (other than what has
	already been described), or juxtaposed mixing was used. For public
	presentation, cross-fading between selective event segments was employed to
	represent as much sonic and behavioral diversity as possible within a
	reasonable listening time frame.
\end{enumerate}

\section{Listening to Immersive Audio}
\label{app:Listening}

We use headphones (Sony MDR-V6) to listen to the binaural renderings and a
custom surround system to listen to the Hexagonal F 6.1 surround renderings.
The latter consists of a subwoofer and six co-planar satellite speakers
(located at head level and centered around the sweet spot where the listener
sits).

Headphone listening is convenient and captures much of the spatial audio.
However, Hexagonal F surround presents a richer sound environment with much
improved directivity. The latter listening configuration is often necessary for
discriminating sources in complex acoustic scenes and low-level or distant
acoustic sources, as described in discussed in the humpback vocal harmonizing.

\section{Surface observation}

For completeness, we note that both video and photographic recording of surface
behaviors were used simultaneously during the HAP recording sessions. The
camcorder (Sony FX30 Digital Cinema Camera) and digital still camera (Nikon Z6
Mirrorless Camera, 400mm telephoto lens with teleconverter) were synchronized
over a wireless link to the time-code generated by the digital audio recorder
for the four HAP signal channels. Thus, surface observations and undersea
acoustics could be accurately cross-referenced.

\end{document}